\documentclass{singlecol-new}

\usepackage{graphicx}

\usepackage[square,sort]{natbib}
\def\newblock{\hskip .11em plus .33em minus .07em}
\makeatletter
\newdimen\mathindent
\def\equation{\@beginparpenalty\predisplaypenalty
  \@endparpenalty\postdisplaypenalty
  \refstepcounter{equation}\trivlist \item[]\leavevmode
  \hbox to\linewidth\bgroup $\@eqnnum\m@th\displaystyle\hfil}

\def\endequation{$\hfil\displaywidth\linewidth\egroup\hskip-12pt\endtrivlist}

\makeatother

\begin{document}

\setcounter{page}{1}

\LRH{D. Carpentier, E. Orignac, G. Paulin and T. Roscilde}

\RRH{Disorder in low dimensions}

\VOL{x}

\ISSUE{x}

\BottomCatch

\PAGES{xxxx}

\CLline

\PUBYEAR{2008}

\subtitle{}

\title{Disorder in low dimensions : localisation effects in spin glass wires and cold atoms}

\authorA{D. Carpentier*, E. Orignac, G. Paulin, T. Roscilde}

\affA{Laboratoire de Physique,\\ Ecole Normale Sup\'erieure de Lyon, \\
46, All\'ee d'Italie, 69007 Lyon, France
\newline $^{*}$Corresponding author : David.Carpentier@ens-lyon.fr}

\begin{abstract}
In this paper, we review the recent activity of our group on the study
of disorder effects on  systems displaying phase coherence. These studies have
focused on both the electronic transport through mesoscopic metallic spin
glasses, and  cold atomic gases trapped in a disordered potential.
\end{abstract}

\KEYWORD{Weak localisation; Spin Glass; Atomic gases.}

\begin{bio}
{\bf D. Carpentier} received his PhD in Physics from the University Pierre et
Marie Curie in Paris. He is a CNRS researcher at the Laboratoire de Physique de
l'ENSL. {\bf E. Orignac} received his PhD in Physics from the University
Paris-Sud. He is a CNRS researcher at the Laboratoire de Physique de
l'ENSL. {\bf T.
Roscilde} received his PhD in Physics from the University  of Florence, Italy.
He is an associate professor  at the Laboratoire de Physique de l'ENSL. {\bf G.
Paulin} is a graduate student  at the Laboratoire de Physique de l'ENSL.
\end{bio}

\maketitle

\section{Introduction}

The last ten years have witnessed tremendous developments in the control of
systems at the nanoscale, allowing the fabrication of artificial structures
that are effectively two-dimensional    \cite{geim_graphene,castroneto2007},
one-dimensional   \cite{ijima_nanotubes_synthesis,charlier_nanotubes} or
zero-dimensional \cite{hanson_dots}. Simultaneously, progress in optical
trapping of atoms have also lead to the possibility of realizing quantum
systems of controllable effective dimensionality \cite{bloch08_manybody}. From
the theoretical point of view, these low dimensional systems are expected to
exhibit a rich physics resulting from quantum coherence which is now becoming
experimentally accessible. A particularly striking property in low dimension is
the phenomenon of Anderson localisation \cite{anderson_localisation}, in which
disorder makes the wavefunction of particles localised. In both
 one \cite{mott_loc} and two \cite{abrahams_loc}
dimensions, any small amount of disorder is expected to localise all particles.
In one dimension, random matrix theories have provided a highly detailed
understanding of the localisation phenomenon  \cite{beenakker_rmt}, while in
two dimensions the regime of weak localisation has been thoroughly studied. Our
group has been exploring at a theoretical level the physics of localisation in
low-dimensional artificial structures. Two main lines of research have been
emphasised. In the first one, \cite{carpentier:2008,paulin1}, reviewed in
Sec.~\ref{sec:spinglasses},  mesoscopic samples of metallic spin glasses are
analysed. In these systems, the underlying disorder is generated by the spin
glass state, and weak
 localisation is used as a tool to probe its
statistical properties. The
fluctuations of observable quantities such as the conductance
and the correlations between them in the weak
localisation regime are shown to provide information on  the overlap between
the states of the spin glass of the spin glass state.
In the second line of
research, reviewed in Sec.~\ref{sec:coldatoms},
trapped bosonic atomic gases are
considered. \cite{Dengetal08,Roscilde08,Roscilde09}
In these systems, in contrast to the spin glass systems, there is a
high degree of control of  the disorder acting on the system, and
the aim is to  explore how Anderson localisation is affected by the
interaction and the statistics of the quantum
particles, a long standing problem of condensed matter physics. \cite{fisher_boson_loc}


\section{Electronic transport through a spin glass nanowire}
\label{sec:spinglasses}

\subsection{Short overview of spin glass physics}

 Spin glasses are amorphous magnetic phases. The canonical examples of
such glasses are obtained by doping with magnetic impurities noble non-magnetic
metals, such as Cu:Mn, Ag:Mn, Au:Fe. At high enough concentrations, the coupling
between the randomly located impurities lead to the appearance of a new
magnetic phase. In this phase, the spins are frozen but lack any simple long range
order. The analogy with conventional glasses is at the origin of the denomination
of this magnetic phase. The understanding of the nature of this frozen phase and the associated
transition from the paramagnetic phase has motivated numerous experimental,
theoretical and numerical works over the last three decades. Experimentally, this
freezing manifest itself through the sharp reduction of magnetic susceptibility $\chi'(\omega\to 0,T)$
below the glass
transition (but not on {\it e.g} the specific heat). This decrease of $\chi'(\omega\to 0)$ is accompanied
 by a broad spectrum of the imaginary part of $\chi''(\omega)$, manifesting the
 very slow magnetic relaxation in the spin glass phase. Indeed, no intrinsic time scale characteristic of this relaxation
can be determined experimentally : the spin glass is said to be aging. Its only relevant time scale
 is its age $t_{w}$ :
correlation and response functions are not stationary and depend explicitly on both times $t_{w}$ and $t_{w}+t$.
This dependence on history of the spin glass properties possesses
remarkable properties : for example, the spin susceptibility $\chi'(\omega\to 0,T)$
crucially depends on the presence or not of a small magnetic field upon cooling through the
transition temperature (field cooled or zero field cooled protocols). The response of the
aging properties of the thermo-remanent magnetisation to small variations of temperatures
manifest the so-called  memory and rejuvenation effects.

 These unusual properties have focused the attention of a large community of condensed matter
 theoreticians. Most studies have focused on the Ising spin glass, expected to describe the low energy excitations
in the presence of  spin anisotropy (originating from {\it e.g.} spin-orbit coupling).
 For Ising spin glasses, G. Parisi proposed a mean field Ansatz which is now understood to be exact
 \cite{Parisi:1980,Guerra:2002,Talagrand:2006}. In this mean field solution for the statics,
the phase space is complex with different ergodic components (see  \cite{Mezard:1987,Parisi:2007}
for pedagogical reviews). These ergodic components correspond
to different pure states. Naively, we expect from such a mean field scenario
that whenever we enter the spin glass phase by crossing the transition temperature,  we end
up in a randomly chosen ergodic component where relaxation dynamics takes place. If we repeat the quenching
procedure in the same sample, we can end up in another of these ergodic component. History
dependence follows directly. The number of different ergodic components (or ``energy valleys'')
depends of the complexity
of the phase space, and is encoded in the distribution of probability of distances between these
pure states. These distances are conveniently encoded using the overlap between
spin configurations. Let us consider two classical spin configurations
$\{\vec{S}^{(1)}_{i}\}_{i}$ and $\{ \vec{S}^{(2)}_{i}\}_{i} $ (same positions of the spins, but different orientations).
The overlap between these configurations is defined as
\begin{equation}
\label{eq:overlap}
Q_{12} = \frac{1}{N_{imp}} \sum_{i=1}^{N_{imp}}
\vec{S}^{(1)}_{i}. \vec{S}^{(2)}_{i} ,
\end{equation}
 where $N_{imp}$ is the number of magnetic impurities (spins). For Ising spins, this overlap will count by how many
 spins flips the configurations $\{\vec{S}^{(1)}_{i}\}_{i}$ and $\{ \vec{S}^{(2)}_{i}\}_{i} $ differ.
  The distribution of overlaps is a central
 object in the Parisi's mean field solution  \cite{Mezard:1987,Parisi:2007},
 and its distribution is indeed the proposed order parameter for the spin glass
 transition. Moreover besides allowing to encode part of the complexity of the phase space, this
 overlap is also a useful tool to measure variations of a spin configuration under variation of an
 external parameter (temperature or magnetic field), or as time evolves, for which configurations
 $1$ and $2$ corresponds to different times (see {\it e.g}  \cite{Cugliandolo:2002}).

 Coming back to the mean field solution, its relevance to the physical situation $D=3$ is still an
 active domain of research. Besides this static solution, solutions of mean-field dynamics have
 been obtained by Cugliandolo and Kurchan (see  \cite{Cugliandolo:2002} for a recent detailed review).
 Mean field aging dynamics can also be described in the so-called weak-ergodicity scenario
 proposed by Bouchaud   \cite{Bouchaud:1992}.
 Moreover, even the complexity of the phase space have been questioned. In another proposal
 based on scaling ideas \cite{Fisher:1986,Bray:1986}, the peculiar properties of spin glasses are
 attributed solely to the aging dynamics, without any need for a complex phase space.
  In this phenomenological theory, the slow dynamics is described
 analogously to the evolution of a quenched ferromagnet : domains of various sizes and shapes evolve
 slowly in time. It is fair to say that, while these numerous results have deepened our understanding
 of the amazing properties of spin glasses, a complete understanding of real spin glasses is
 still lacking. The purpose of our work is to propose another experimentally accessible probe of metallic
 spin glasses, allowing interesting and crucial comparisons with current theories.

\subsection{What does coherent electronic transport probe in a spin glass ?}

 In this work, we take advantage of the metallic nature of canonical spin glasses and consider their
 electronic transport properties in the coherent regime. In metals, electron's inelastic scattering
 off {\it e.g.} phonons limit the phase coherence of these electrons.  This dephasing can be described
 phenomenologically by  a dephasing length scale $L_{\phi}(T)$ beyond which
 phase coherence can be safely neglected when describing transport properties. At room temperatures,
 the high density of phonons induce an $L_{\phi}(T)$ much smaller than conductors sizes : the classical
 Drude theory correctly describes  the conductance of metallic wires. However upon lowering the
 temperature, $L_{\phi}(T)$ increases up to a few $\mu m$, allowing to study transport in wires
 of size comparable to $L_{\phi}(T)$. When describing the electronic transport through such devices,
 interferences effects have to be taken into account : this is the so-called mesoscopic regime
  (for a recent and very detailed introduction to this field, see  \cite{Akkermans:2007}). In this regime,
 the conductance is sample dependent. Moreover,  when applying a small magnetic field transverse to the wire,
 the conductance fluctuates in a random manner. The corresponding magneto-conductance trace is
 considered as a fingerprint of the configuration of disorder in the sample : two samples with the same
 amount of disorder  differ by the exact realization of the random potential. Hence they display
 magneto-conductance traces   that are different, but with the same amplitude.

    In a metallic spin glass, part of the randomness consists in the random orientations of the impurities magnetic
moments. The influence of these random spins on the fluctuations of the
conductance was soon realised by Al'tshuler and Spivak \cite{Altshuler:1985}
and specific consequences of the droplet theory of spin glasses were discussed
in  \cite{Feng:1987}. Experimentally, de Vegvar {\it et al.} pioneered the
 mesoscopic transport studies in metallic spin glasses by clearly demonstrating the existence of reproducible
 magneto-conductance traces below the glass transition in Cu:Mn wires \cite{deVegvar:1991}. By using
 the antisymmetric component of conductance upon time-reversal symmetry, they attributed the appearance
 of these coherent conductance fluctuations  to the freezing of impurities spins. In parallel, Israeloff  {\it et al.}
 focused on the electrical noise in spin glass Cu:Mn wires, {\it i.e.} on the time dependence of conductance
 fluctuations \cite{Israeloff:1989,Alers:1992,Meyer:1995}.
 The dynamical characteristics of these spontaneous fluctuations were analysed
 in view of both the mean-field and the droplets theories of spin glasses (see  \cite{Weissman:1993} for a review).
  However, comparison with existing theoretical ideas   \cite{Altshuler:1985,Feng:1987} was hampered
  by the necessity of averaging over the conductance distribution in theoretical predictions. Later,
  universal conductance fluctuations were analysed in diluted magnetic semiconductor in the spin glass
  regime  \cite{Jaroszynski:1998}. In these compound, a rapid increase of the amplitude of these fluctuations
  was noted below the spin glass transition temperature, as well as the usual field cooled/zero field cooled
dependence of these fluctuations. More recently, similar studies have been conducted on Au:Fe
wires  \cite{Neuttiens:1998,Neuttiens:2000}. These different studies have validated the use of mesoscopic
conductance fluctuations as an interesting new tool to probe the spin glass freezing. Most of this work
have focused on either the amplitude of these fluctuations, or their time-dependence (electrical noise).
In the present study, we consider the correlation between traces of magneto-conductance as a new and
most relevant way to analyse spin glass properties.

\subsection{Probing the distribution of overlaps through conductance correlations}

We now focus on the description of electron's coherent transport in  a metallic spin glass. In this
paper we will focus on the spin glass phase, and motivated by previous experimental results we will assume that
the dephasing length scale $L_{\phi}(T)$ (which is dominated by spins able to flip during the diffusion time scale)
is comparable with sample size. The spins that do not contribute to $L_{\phi}(T)$ are then described by
frozen classical spins with random orientations (we do not assume these spins to be Ising like).
We consider lattice model describing the electrons in the presence of these impurities which contribute both to a
scalar potential $v_{i}$ and a magnetic part $ \vec{S}_i $ :
\begin{equation}
\label{def:H}
H = \sum_{<i,j>, s} t ~c^\dagger_{i,s} c_{j,s}
+ \sum_{i,s} v_{i} ~c^\dagger_{i,s} c_{i,s}
+ \sum_i J ~\vec{\sigma}_{ss'} . \vec{S}_i  c^\dagger_{i,s} c_{i,s'}
\end{equation}
 The transport properties of this model in the coherent regime were analysed both using diagrammatic  perturbation theory \cite{Altshuler:1985,carpentier:2008}, and a numerical Landauer approach \cite{paulin1}.
  This last technique consists in obtaining the conductance of a finite sample for a given configuration of
disorder, {\it i.e} without any averaging procedure. Following Landauer formula, this conductance is deduced
from the electron's scattering matrix of the system connected to two reservoirs. In this numerical procedure,
the scalar site-disorder $v_{i}$ are uniformly distributed in $v_i \in [-W/2, +W/2]$ in units of $t=1$. We typically used $W$ of order $0.5$. The spin glass configuration is mimicked numerically by choosing
spins $\vec{S}_i$ randomly on the unit sphere, without any spatial correlations. As explained below special
care was devoted to correlations (overlap) between different spin configurations.

\begin{figure}[!h]
\caption{Function $F(x=L/L_{\phi})$ describing the scaling of the variance of
the conductance distribution without magnetic impurities : see eq.
(\ref{eq:F_x}). The function $H(x)$ represented on the right of the figure,
describes the scaling of this variance when magnetic impurities are present,
and a new dephasing length $L_{m}$ appears : see eq. (\ref{eq:G}). }
\centerline{
\includegraphics[width=6cm]{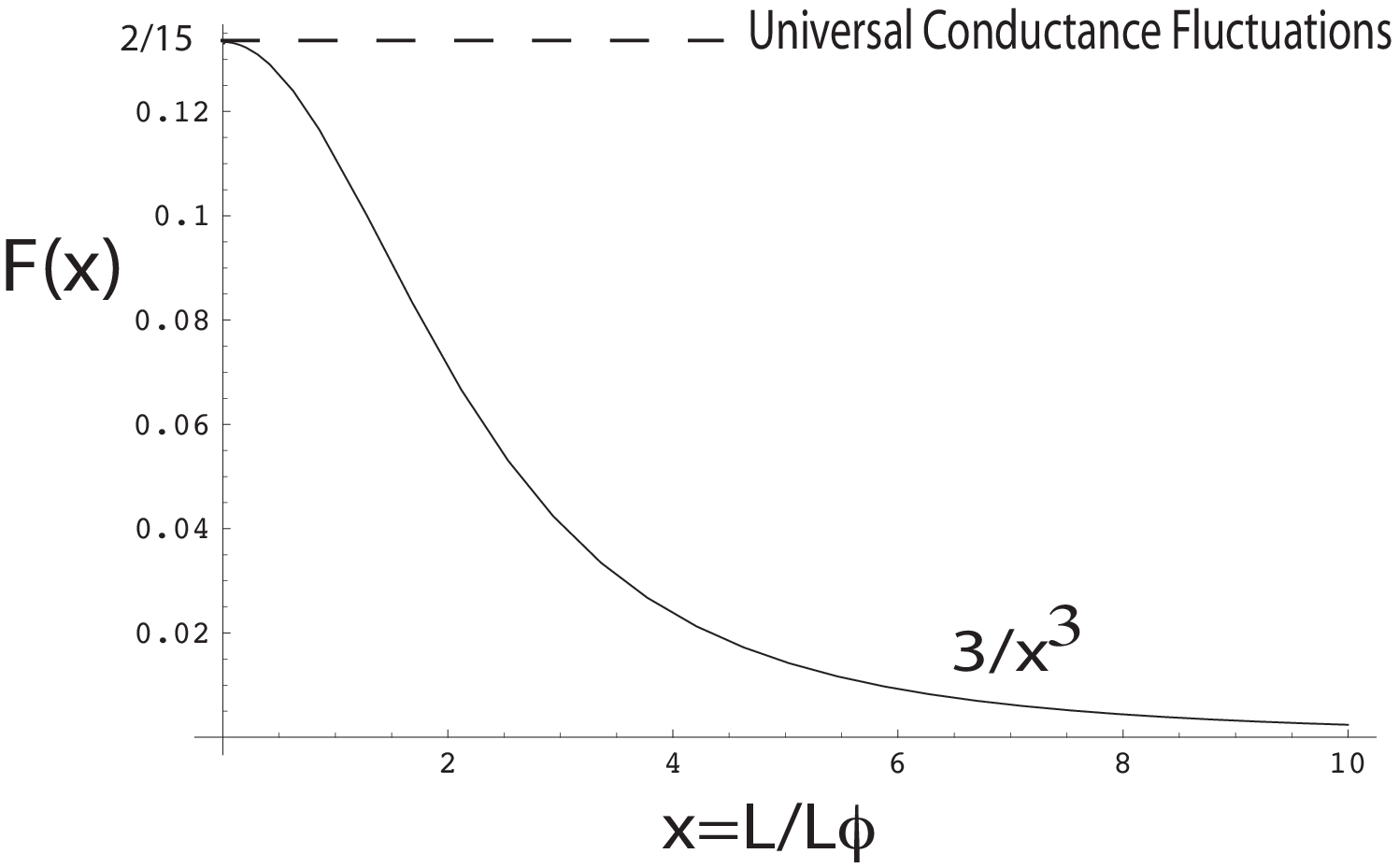}
\includegraphics[width=6cm]{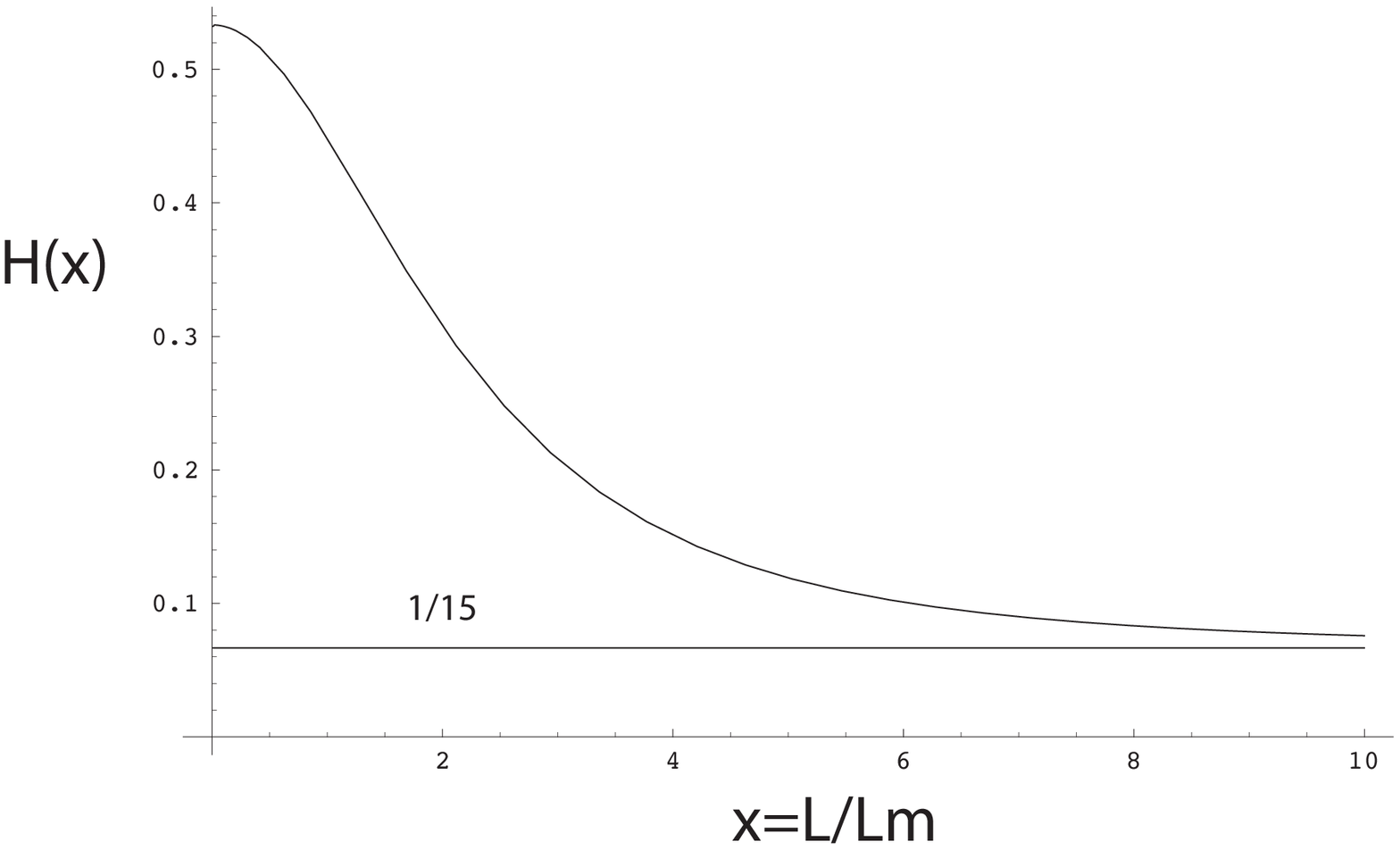}
} \label{fig:Fx}
\end{figure}

Let us first review the situation without magnetic disorder ($J=0$). For a nanowire of size $L_{y},L_{z} \ll L_{\phi} \simeq L_{x}$, coherent electronic diffusion is  effectively
one-dimensional (1D). In this quasi-1D regime, the Anderson's  localisation length scales proportionally to the
number of propagating modes $N_{e}$ : $\xi = N_{e} l_{e}$ where $l_{e}$ is the elastic mean free path.
 For wire's length $L_x$ much smaller than $\xi$, in the so-called weak-localisation regime, the distribution of
conductance $P(G)$ is Gaussian. Without magnetic impurities ($J=0$), its variance
reads
 $\left< (\delta G )^2 \right>_{V}  = s^2 F (L/L_{\phi})$ with  \cite{Pascaud:1998} (see Fig. \ref{fig:Fx})
\begin{equation}\label{eq:F_x}
 F (x)  =
 12  \left( \frac{e^2}{h} \right)^2
\sum_{n=1}^{\infty}
\frac{1}{[n^2 \pi^2  +  x^{2}]^{2} }
 = \frac{3}{x^4}
\left( x^2 ~\textrm{csch}^2 x +x \coth (x)  - 2 \right) ,
\end{equation}
where $\langle  \rangle_{V}$ denotes an average over the random scattering potential
$V({\bf r})=\sum_{i}v_{i}\delta({\bf r}-{\bf r}_{i})$
, and
$\delta G $ is defined relative to the average conductance $  \delta G = G  -\langle G \rangle_{V} $.
In this formula $s$ is the level degeneracy, which corresponds to $s=2$ for spins $1/2$ .
 For an infinite dephasing length scale ($x=0$), we recover the universal conductance value
 $\left< (\delta G )^2 \right>_{V} = s^2 \frac{2}{15}$, whereas $\left< (\delta G )^2 \right>_{V}$
 vanishes as expected in the classical limit $L\gg L_{\phi}$.

 In the presence of magnetic impurities, the conductance $G$ is now a function of both the configuration
of scalar potential $V({\bf r})=\sum_{i}v_{i}\delta({\bf r}-{\bf r}_{i})$ and the configuration of
frozen spins $\{S_{i}\}$. For a given configuration of spin $\{S_{i}\}$, we can consider the distribution
of $G(V,\{S_{i}\})$ as $V$ varies. This distribution $P_{V}(G(\{S_{i}\}))$ is still gaussian and independent of the configuration of spins. The
length dependence of its variance is modified by the presence of the symmetry breaking disorder.
 A new magnetic dephasing length $L_{m}$ appears (here given to order $J^{-1}$) :
 \begin{equation}
 L_{m}^2 = \hbar D \left( 4\pi n_{\rm imp} \rho_{0} J^2 S^2 \right)^{-1},
 \end{equation}
  where $D$ is the diffusion coefficient in the sample, $n_{\rm imp}$ the density of impurities and
  $\rho_{0}$ the electron's density of states. In the limit $L_{m},L \ll L_{\phi}$, the variance of $P_{V}(G)$ reads
 \begin{align}\label{eq:G}
\left< \left[\delta G (\{S_{i}\})\right]^2 \right>_{V}  & = H(L/L_{m})\\ \nonumber
&=
\frac14 F \left(0 \right)
+ \frac34 F\left(\frac{2}{\sqrt{3}}\frac{L}{L_{m}}\right)
+ \frac14 F\left(\sqrt{2} \frac{L}{L_{m}}\right)
+ \frac34 F\left( \sqrt{ \frac{2}{3} }    \frac{L}{L_{m}}\right)   .
\end{align}
  This function $H(x)$ is represented in figure \ref{fig:Fx} for $L_{\phi} \to \infty$.

\begin{figure}[!h]
\caption{Distribution of the variation of conductance between two spin
configurations $\{S_{j}^{(1)} \} $ and $\{S_{j}^{(2)} \}$ in a given sample
$V$. These distributions are obtained using a numerical Landauer approach for
the model (\ref{def:H})  \cite{paulin1}. The coupling $J$ between the spins of
the electron and the impurities was chosen as $J/t=0.1$, the system size was
$40\times1600$. For each configuration of spins, $5000$ configurations of
scalar disorder $V$ were generated. The histograms over these $5000$ points are
shown for $5$ pairs of spin configurations.
 While the distributions for pairs with a spin overlap $Q\simeq 0.91$ collapse within a good approximation,
the distribution corresponding to $Q=0.99$ clearly deviates from these other
distributions. This result illustrates the parameterisation of the distribution
of $G  \left(V,\{S_{j}^{(1)} \}\right)  - G\left(V,\{S_{j}^{(2)} \}\right)$ by
the spin overlap $Q$.}
\centerline{
\includegraphics[width=10cm]{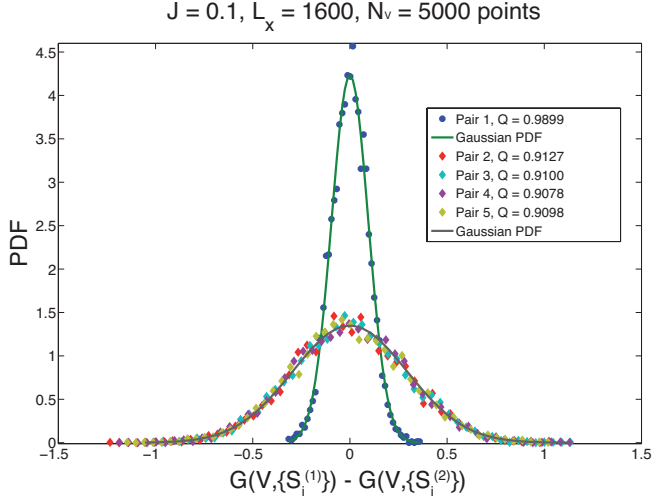}
} \label{fig:diffG}
\end{figure}

 We now focus on the dependence of this conductance on variations of the configurations of spins. For a given configuration of $V$, we consider the variation of conductance between two spin configurations (which
 differ by the orientation of the frozen spins $\{S_{i}\}$, but not their position) :
\begin{equation}\label{eq:diffG}
G  \left(V,\{S_{j}^{(1)} \}\right)  - G\left(V,\{S_{j}^{(2)} \}\right)  .
\end{equation}
As $V$ is varied, this quantity fluctuates similarly to $G$. The corresponding distribution, obtained using the
numerical Landauer method, is represented  in Fig.~\ref{fig:diffG}
   for several pairs of configurations $\{S_{j}^{(1)} \} $ and $\{S_{j}^{(2)} \}$.
 As expected, this distribution is still gaussian, with a vanishing average. For a fixed geometry, its variance
 appears to be parameterised by the overlap (\ref{eq:overlap}) between the two spin configurations : the
 distributions of $G$ for different pairs of configurations with similar overlap collapse.
  Indeed, its variance is entirely encoded by the average correlation between the conductances in two different frozen spin configurations
 $,\{S_{j}^{(1)} \}$ and $\{S_{j}^{(2)} \}$ :
\begin{equation}\label{def-GG}
(\Delta G)^2_{S^{(1)}S^{(2)}} =
\left< \delta  G  \left(V,\{S_{j}^{(1)} \}\right)  \delta G\left(V,\{S_{j}^{(2)} \}\right) \right>_{V},
\end{equation}
 where
 $\delta  G  (V,\{S_{j}^{(1)} \})  = G  (V,\{S_{j}^{(1)} \})  - \langle G  (V,\{S_{j}^{(1)} \}) \rangle_{V}$.
  The length dependence of these correlations is obtained by extending (\ref{eq:G})
 \citep{Altshuler:1985,carpentier:2008}.
 In the same limit $L_{m} \ll L_{\phi}$ as in (\ref{eq:G}) , it reads
\begin{align}\label{eq:DeltaG}
(\Delta G)^2_{S^{(1)}S^{(2)}}  & =
  \frac14 F \left(\sqrt{1-Q}\frac{L}{L_{m}}\right)
+ \frac34 F\left(\sqrt{1+\frac{Q}{3}}\frac{L}{L_{m}}\right)    \nonumber \\
&+ \frac14 F\left(\sqrt{1+Q} \frac{L}{L_{m}}\right)
+ \frac34 F\left( \sqrt{1-\frac{Q}{3}}    \frac{L}{L_{m}}\right)
\end{align}
where  $Q$ is the overlap between the two spin configurations defined in (\ref{eq:overlap}).
 This result is the starting point for our proposal to measure the overlap between random
spin configurations : in a given sample when orientations of the spins are varied, the variations
of the quantity (\ref{eq:diffG}) are entirely encoded by the overlap between the corresponding
spin configuration. Such a measurement would allow for the first experimental determination of
spin configuration's overlap.

\begin{figure}[!h]
\caption{Magnetoconductance traces obtained by a numerical Landauer approach
\cite{paulin1}. The different traces correspond to the same sample, {\it i.e}
configuration of scalar disorder, but different configurations ({\it i.e}
orientations)
 of the frozen spins. }
\centerline{\includegraphics[width=7cm]{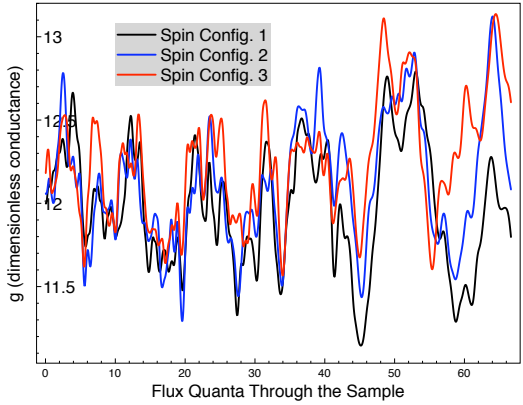}}
\label{fig:Traces}
\end{figure}

 A crucial step needed to experimentally measure (\ref{eq:diffG}) is to sample the corresponding
distribution. This distribution was defined by considering the statistics of
$G$ when the potential $V$ is varied. However, one cannot vary the diffusing
potential in a semiconducting spin glass without modifying completely the spin
glass configuration ! We thus have to resort to the usual ergodic hypothesis of
weak localisation (see  \cite{Tsyplyatyev:2003} for a recent discussion).
 Following this hypothesis, the statistics of $G$ are usually determined experimentally by
considering a magneto-conductance trace of a given metallic sample. When a flux is applied perpendicular
to the sample, the conductance fluctuates in a random but reproducible way (see Fig.~\ref{fig:Traces}).
The variance of the
corresponding distribution approximates correctly the variance of the distribution $P_{V}(G)$  \cite{Tsyplyatyev:2003}.

\begin{figure}[!h]
\caption{Proposed protocols to measure the correlations (overlap) between spin
configurations corresponding to different quenches or different waiting times.
In the first case, the different spins states $\{\vec{S}_{i}^{(n)}\}$ are
labelled by the index of the quench $n$. All these states correspond to the
spin configuration obtained after a time $t_{w}$ after entering the spin glass
phase. At this time $t_{w}$, the spin configuration is sampled through the
measurement of the corresponding magnetoconductance trace. The correlation
between these traces gives access to the overlap between the corresponding spin
states. In the second case, the spin configuration correspond to different
waiting time $t_{w}^{(n)}$. Similarly, these states are probed through
magnetoconductance measurements. }
 \centerline{
\includegraphics[width=6cm]{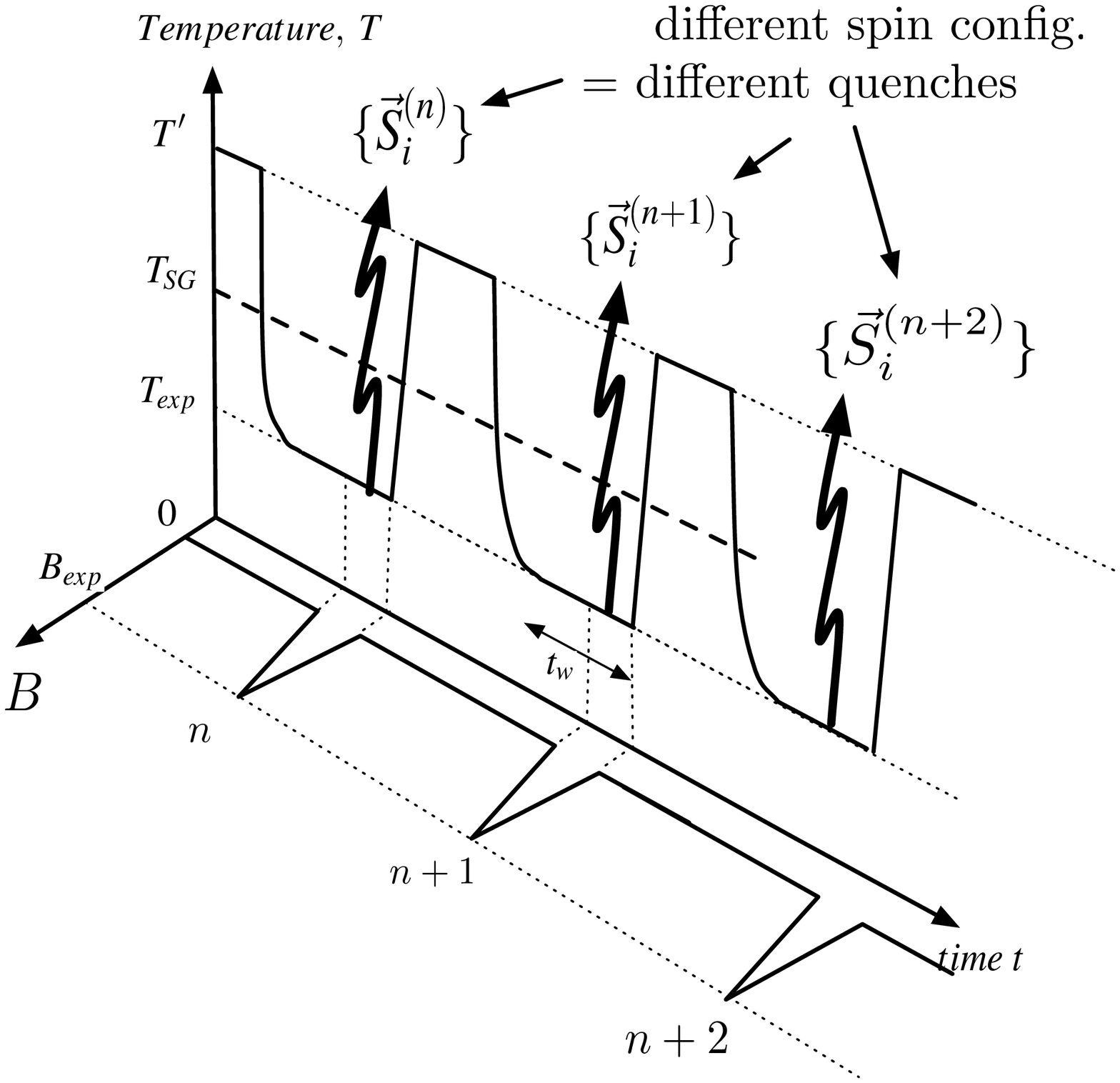}
\includegraphics[width=6cm]{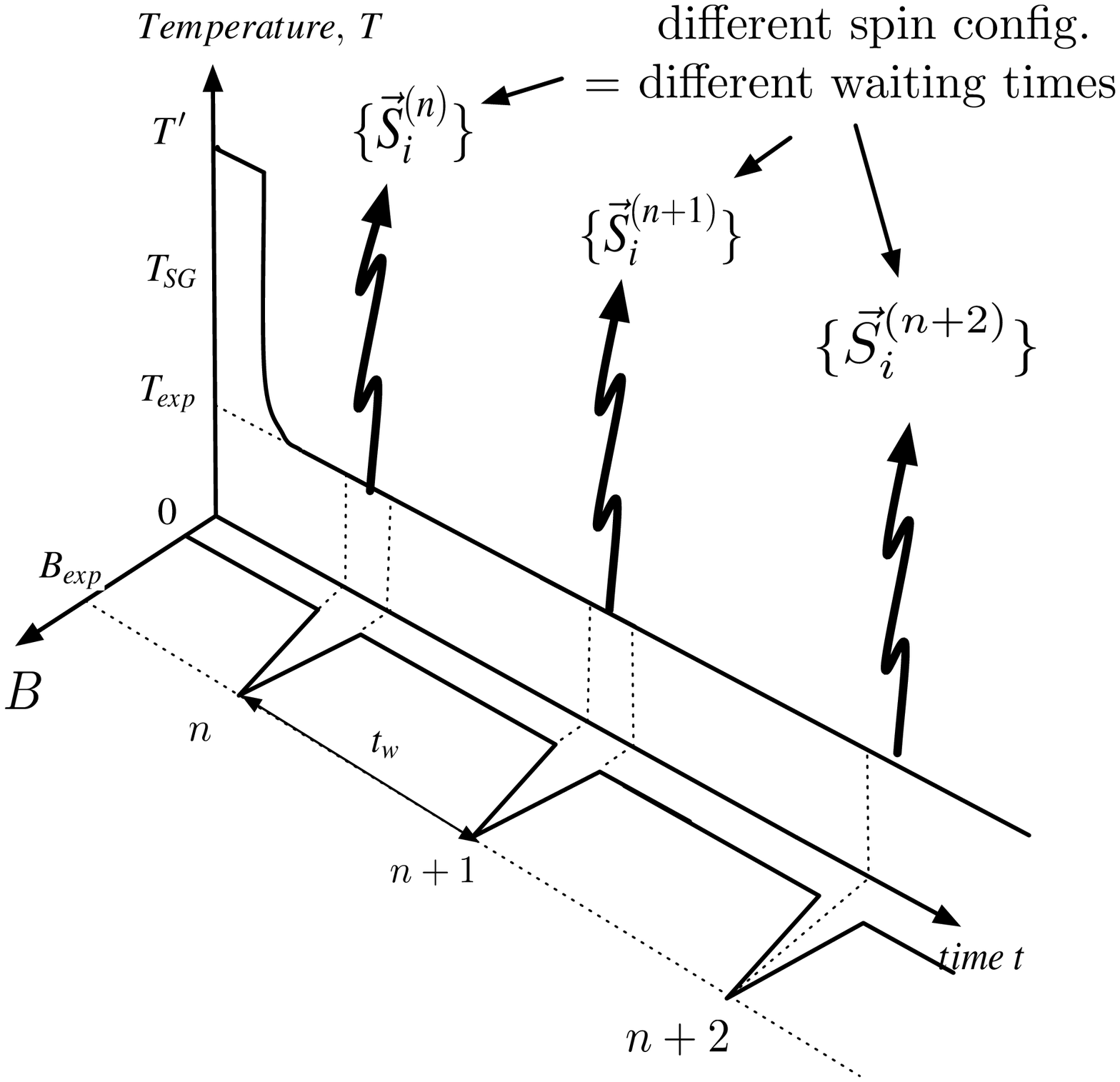}
} \label{fig:Protocols}
\end{figure}
 Following this line of reasoning, the proposed quantity to measure instead of (\ref{def-GG}) is
\begin{multline}\label{eq:Bsampling}
\langle \delta G  (V,\{S_{j}^{(1)}\})  \delta G(V,\{S_{j}^{(2)}\})\rangle_{B} = \\
\frac{1}{B_{\rm max}-B_{\phi}} \int_{B_{\phi}}^{B_{\rm max}}
 \delta G  (V,\{S_{j}^{(1)}\},B)  \delta G(V,\{S_{j}^{(2)}\},B) dB
\end{multline}
A crucial approximation made in using (\ref{eq:Bsampling}) as an evaluation of
(\ref{def-GG}) is the neglect of the spin glass response to the (weak) magnetic
field. A crucial step in this possible experimental route would thus consist in
determining the optimal maximum field $B_{\rm max}$ below which this spin glass
response can be neglected, but high enough to provide the necessary sampling of
the statistics of $G$.

In conclusion, we have proposed a first route to measurements of spin
configurations overlap in metallic mesoscopic conductors. This route consist in
determining the magneto-conductance traces for different spin configurations.
The correlation between these traces gives access through eq.~(\ref{eq:DeltaG})
to the overlap between the corresponding spin configurations.  In a spin glass,
these different spin configurations can be accessed either by successive
cooling below the glass transition, or {\it e.g} by different waiting times.
The proposed protocols corresponding to these two situations are illustrated in
Fig.~\ref{fig:Protocols}.

%



\section{Anderson localisation with cold atoms}
\label{sec:coldatoms}
\subsection{Introduction}

The field of ultracold atoms has emerged in recent years as a new research
field at the frontier between atomic, molecular and optical physics and
condensed matter physics.~
\cite{jaksch05_coldatoms,lewenstein07_coldatoms_review,bloch08_manybody}.
Trapping potentials of arbitrary shapes can be Fourier-synthesised via the
superposition of different laser standing waves. In particular a single
standing wave confines the atoms into planes, with negligible inter-plane
coupling (over the time duration of the experiments, typically 100ms-1s); two
orthogonal standing waves confine the atoms into tubes, while three standing
waves create a cubic optical lattice. More sophisticated lattice geometries can
be then achieved by superposing further spatial Fourier components. Hence
optical potentials give unprecedented access to low-dimensional quantum systems
and to arbitrary lattice Hamiltonians.
 When ultracold atoms are loaded in the lowest band of optical lattices,
 their physics is faithfully described by Bose- and Fermi-Hubbard models.
 The depth of the optical lattice allows to control the inter-site
 hopping, and hence the relevance of interactions
 against the kinetic energy over several orders of magnitude. Interactions are generally short-ranged
 (on-site only), and can be further tuned with the use of Feshbach
resonances~ \cite{roberts98_feshbach_rb}.  Hence experiments can
span all regimes of lattice Hamiltonians, from the weakly interacting
one to the strongly interacting one, and probe fundamental phenomena
such as the onset of Mott insulating physics and the suppression of
quantum coherence  \cite{Greineretal02, Schneideretal08}.

 Yet cold atomic samples contain typically $10^5-10^6$ atoms;
when standing waves are applied to divide the system into planes or
tubes, each layer/tube contains typically   $10^3-10^4$ / 10-100 atoms
respectively.  Hence trapped cold atoms can be
viewed as a particular instance of mesoscopic systems  \cite{huang_meso},
where the finite system size, dictated by a global (parabolic) trapping
imposed on the atoms,  can be far exceeded by the quantum coherence
length. Coherent quantum many-body phenomena out of equilibrium can be
probed in the system via a real-time control on the Hamiltonian parameters,
and they have been demonstrated in recent experiments on Bloch oscillations
in periodic lattices  \cite{Dahanetal96},  Josephson oscillations in a double-well potential  \cite{Albiezetal05},
coherent dipole oscillations in a parabolic trap  \cite{Fertigetal05},  transitions from integrability
to non-integrability  \cite{kinoshita_cradle}, etc.

  Our recent theoretical activities on mesoscopic aspects of trapped cold
  atoms have been focused on the  physics of Anderson
localisation \cite{anderson_localisation} of interacting bosonic atoms.
Anderson localisation of interacting
bosons is a long-standing problem in condensed
matter physics  \cite{fisher_boson_loc}, and various experimental realizations
have been attempted with ${}^4$He in a
porous medium such as Vycor, aerogels or xerogels \cite{chan_vycor_localization}
or with disordered Josephson
junctions \cite{vanoudenaarden_josephson_localization}.
However an experimental test of fundamental theoretical predictions is
still lacking.  Thanks to the flexibility in designing the trapping potential,
cold atomic systems offer the possibility to explore one dimensional
geometries, in which strong localisation effects are dramatic
 \cite{mott_loc} and have been extensively studied
theoretically  \cite{lifshitz_desordonne}. Moreover,
the use of optical lattices and Feshbach resonances allow to study the
interplay of disorder and a continuously tuned interaction.

Two methods to realise a controlled random (or pseudo-random) optical potential
are currently available. One consists of imaging a disordered diffusive plate
onto the atomic sample, thus creating a so-called laser-speckle potential
\cite{lye05_speckle, fort05_speckle1d, clement_disorder_quasi_1D,
clement06_speckles}. The reduction of the speckle size below the bosonic
healing length has recently allowed to clearly observe real-space signatures of
Anderson localisation on the density profile of a highly dilute gas expanding
in a one-dimensional speckle potential  \cite{billy_loc}.
 The second method, to which our recent investigations have been
 devoted, is specific of optical lattice systems, and it is based on
 superposing a dominant (primary) standing wave with a secondary
standing wave with an incommensurate period with respect to the
primary one. In this way a quasi-periodic superlattice is created, whose
physics interpolates between the one of a random system and the one
of a perfectly periodic potential, as we will discuss in the next section.
This approach has also allowed to recently observe clear signatures
of Anderson localisation in expanding one-dimensional atomic clouds
with negligible interaction  \cite{roati_loc}, as well as signatures of
unconventional insulating behaviour in the strongly correlated regime
 \cite{fallani07_boseglass}.

\subsection{Ultracold bosonic atoms  in disordered optical lattices}
\label{sec:disorder-boson}

 Interacting bosons loaded in a deep optical lattice plus
 a random potential are described by a Bose-Hubbard Hamiltonian
  \cite{jaksch_bose_hubbard},
 of the form
\begin{eqnarray} \label{ham}
  H=-J \sum_{\langle ij \rangle} (b^\dagger_{i} b_j + \mathrm{H. c.}) + \frac{U}{2} \sum_i n_i
  (n -1) + \sum_i (\epsilon_i-\mu)  n_i,
\end{eqnarray}
 where $b_i$ is the boson annihilation operator, $n_i=b^\dagger_i b_i$ is the boson
 number operator,
  $J$ is the hopping integral, $U$ is the on-site boson-boson
 repulsion, $i$ runs over the sites of an optical lattice, and $\langle ij \rangle$
 runs over the pairs of  nearest neighbouring sites.
  $\epsilon_i$ is the on-site energy, accounting for the overall trapping
  potential (of the form $V_t (i-i_0)^2$) and for any additional random
  or quasi-random potential.

\begin{figure}
\caption{Schematic phase diagram of the Bose Hubbard model without disorder
(left panel) and with disorder (right panel). MI=Mott insulator, SF=superfluid,
BG=Bose glass, AG=Anderson glass.}
\begin{center}
\includegraphics[width=9cm]{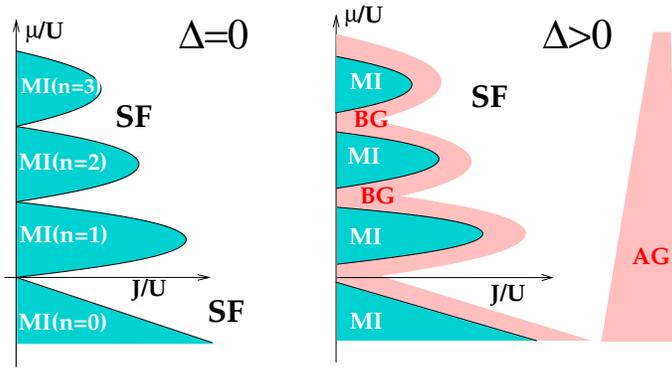}
\end{center}
\label{f.BH}
\end{figure}

  In absence of additional potentials, the Bose-Hubbard model at zero temperature
 features a gapless superfluid (SF) phase with long-range phase coherence for large $J/U$;
 lowering this ratio leads the system to a quantum phase transition towards gapped Mott-insulating (MI)
 phases at integer fillings and with short-range coherence.
 (see Fig.~\ref{f.BH}). Upon introducing a random on-site potential, namely taking $\epsilon_i$ as a
 random variable in the interval $[-\Delta/2, \Delta/2]$, two new phases appear: for large $J/U$,
 an \emph{Anderson glass} (AG) phase, in which weakly interacting particles are Anderson-localised
 by disorder, and their state connects continuously with that of the single-particle localisation
 problem; in the strongly interacting regime (low $J/U$ ratio), a \emph{Bose glass} (BG) phase, characterised
 by Anderson localisation of collective modes of the system, which can be essentially
 pictured as particle-hole excitations involving localised quasi-particle states. Both AG and
 BG are exotic insulators compared to the MI, in that they feature a gapless spectrum and a finite
 compressibility in absence of long-range coherence. The schematic phase diagrams of Fig.~\ref{f.BH}
 can be obtained via mean-field theory  \cite{fisher_boson_loc} and have been repeatedly confirmed via
 numerically exact techniques, such as quantum Monte Carlo (QMC)  \cite{ProkofevS98} and
 density matrix renormalization group (DMRG)  \cite{Rapschetal99}.

\begin{figure}
  \caption{Phase diagrams for the Bose-Hubbard model in a Harper potential. Left panel: phase diagram
  for weak disorder $\Delta/U=0.1$ (in the inset the superfluid fraction and the compressibility gap  are
  plotted as  functions of $t/U$ for $\mu/U=0.25$).
  Right panel: phase diagram for strong
  disorder, $\Delta/U=0.5$.}
\rotatebox[origin=br]{270}{\includegraphics[width=0.33\textwidth]{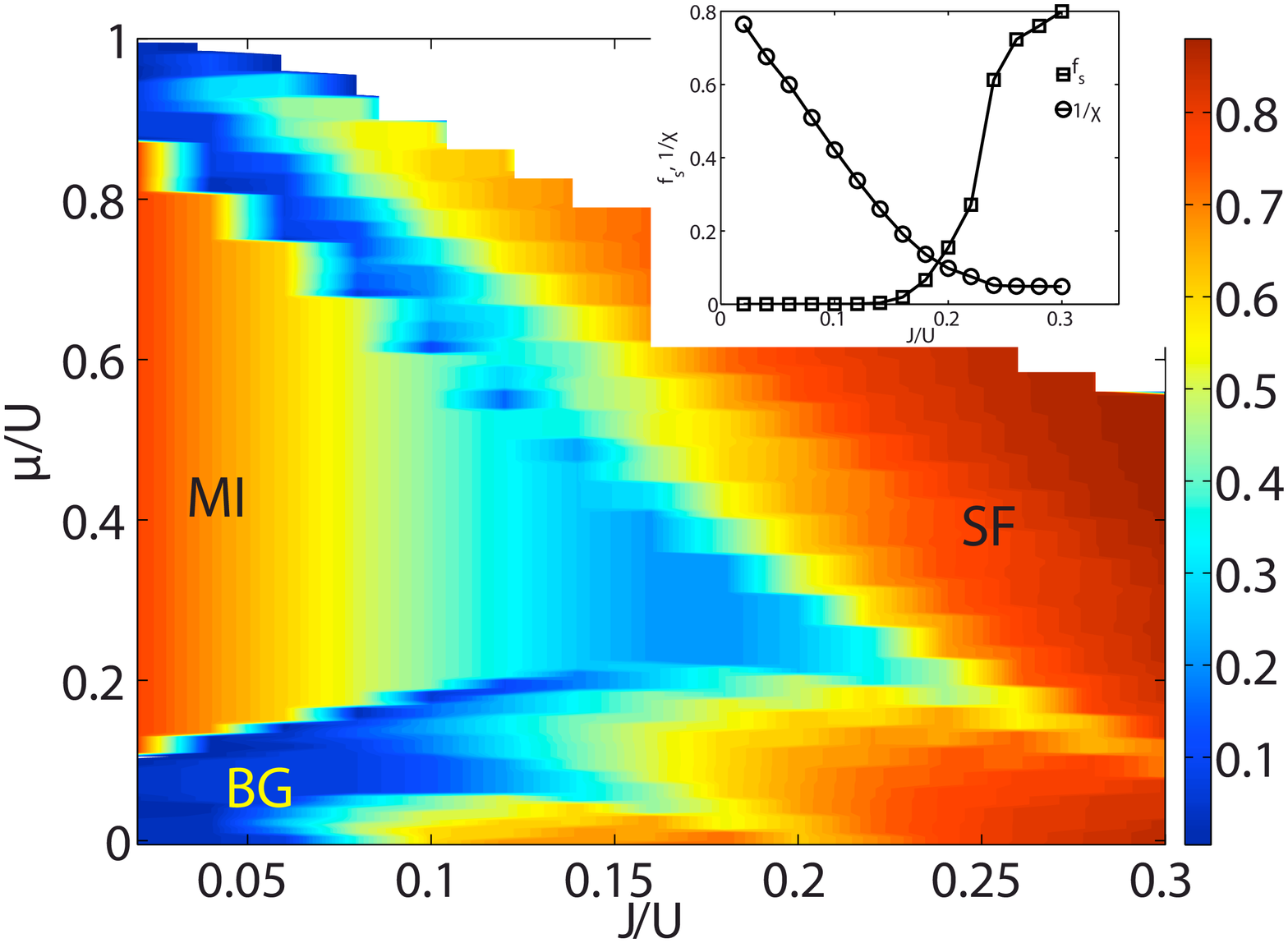}}
\hspace{4mm}
\includegraphics[width=0.51\textwidth]{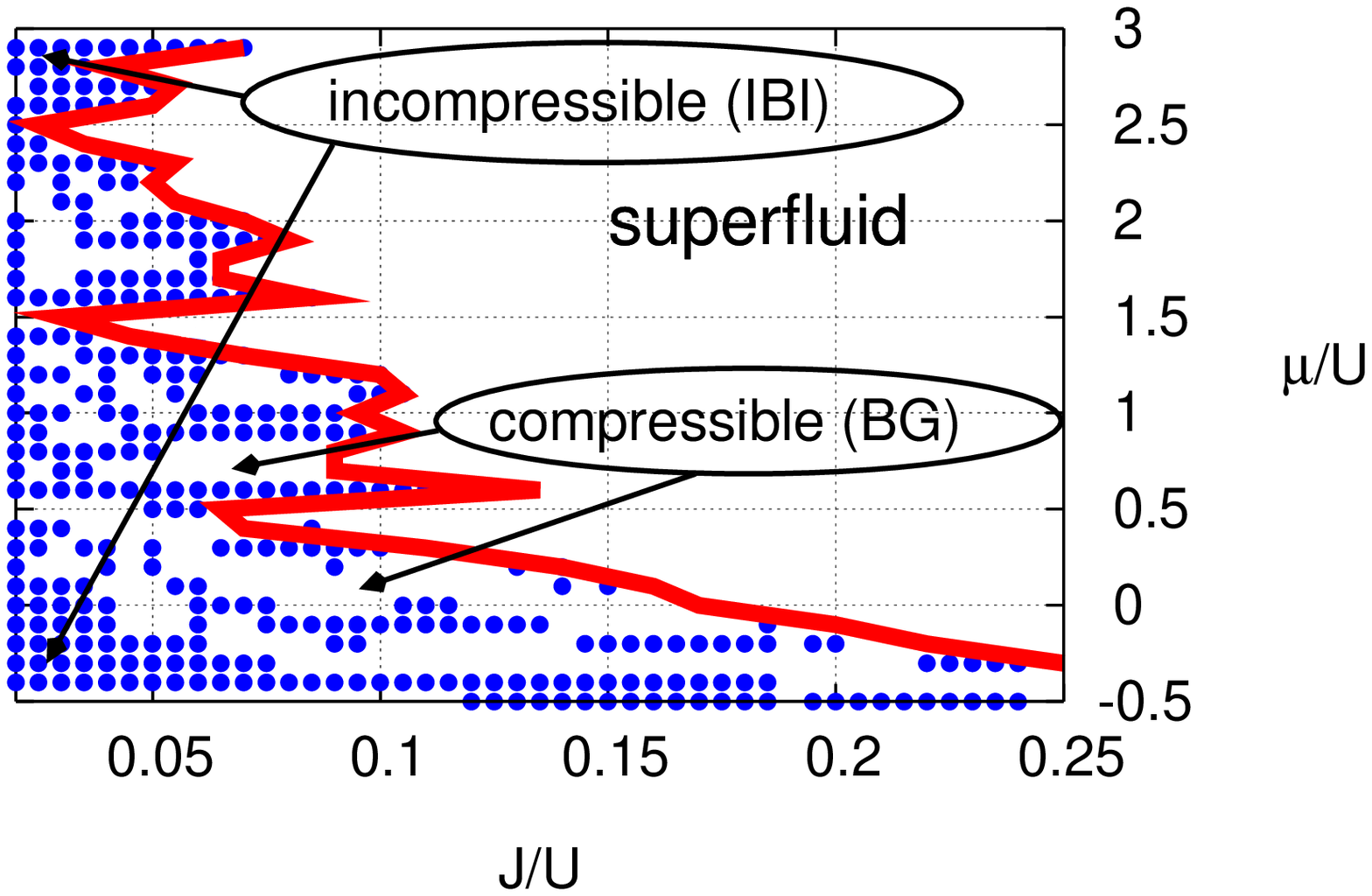}
  \label{f.Harper}
\end{figure}

   In the case of a quasi-periodic superlattice,  the on-site potential takes the
  form of a Harper potential $\epsilon_i=\Delta \cos (2\pi \alpha i + \phi)$ where
  $0<\alpha<1$ is an irrational number and $\phi$ an arbitrary phase factor.
   In the following we will specialise the
  discussion to one-dimensional systems, for which the problem of a single quantum particle
  in the Harper potential has been the subject of extensive investigations in the past \cite{Sokoloff85}.
  A one-dimensional particle in a purely random potential (as the one described
  above) is known to undergo Anderson localisation in all its energy eigenstates
  for any infinitesimal disorder strength, namely the Anderson transition occurs
  at $\Delta=0$. Moreover the single-particle density of states
  is continuous (on average over the disorder realizations). On the contrary a single
  particle in the Harper potential undergoes an Anderson transition from all extended
  to all localised eigenstates for a finite disorder, $\Delta=2J$  \cite{Aubry_harper}.
  The density of states is not continuous (even upon averaging over the arbitrary
  phase $\phi$), and it is structured in bands both on the extended and on the
  localised side of the transition.

    Our investigations have been focused on effect of strong interactions on
   the localisation phenomena of bosons in a Harper potential, following a
   setup which has been recently realised in optical lattice experiments
    \cite{fallani07_boseglass}.  Making use of numerical exact diagonalisation, QMC and
   DMRG methods,
   as well as of analytical bosonization techniques, we have addressed the fundamental
   questions on the topology of the phase diagram in presence of a quasi-random
   potential  \cite{Roscilde08, Dengetal08}. Fig~\ref{f.Harper} shows the phase diagram
   of the Hubbard model in the Harper potential for two values of the disorder potential
   $\Delta/U = 0.1$ (obtained via DMRG) and $\Delta/U=0.5$ (obtained via QMC),
   and for $\alpha=830/1076$, which is chosen as the ratio of the wavelengths
   used in the experimental realization, Ref.~ \cite{fallani07_boseglass}.
   In the case of weaker disorder ( $\Delta/U = 0.1$) one observes the appearance
   of a Bose-glass phase with vanishing superfluid density and finite compressibility
   at the interface between the conventional Mott insulating and superfluid phase.
   Yet, at variance with the case of a truly random potential  \cite{ProkofevS98, Rapschetal99},
   a direct transition is possible between SF and MI close to the tip of the Mott lobe (as shown
   in the inset of Fig.~\ref{f.Harper}). This fact is intimately related with the
   existence of a finite localisation threshold associated with the quasiperiodic
   potential: analogously to what happens at the single particle level, when the intersite hopping
   exceeds a critical value the quasiperiodic
   potential has no localising effect on the collective modes of the system, so that
   the Bose-glass phase is not observed, and a conventional phase
   transition between MI and SF can take place with the onset of gapless and
   propagating collective modes. The single-particle localisation threshold $\Delta=2J$
   is attained for $J/U=0.2$, which is already to the left of the Mott-lobe tip;
   on the other hand interactions play the role of screening the quasi-periodic potential,
   so that even a smaller $J$ is expected to mark the disappearance of the
   Bose glass. This is quantitatively consistent with our data, where the BG
   phase seems to disappear for $J/U\approx 0.15$.

    On the opposite side of the spectrum, a very strong quasi-periodic potential has the
    effect of fully destabilising the MI phase of the homogeneous system at integer fillings;
    indeed for $\Delta/U=0.5$ the amplitude of the fluctuations of the Harper potential
    ($\sim U$) overcome the gap characteristic of the MI phase, which is upper-bounded
    by $U$. Hence the MI phase disappears from the phase diagram, substituted by
    novel insulating phases stabilised by the quasi-periodic potential. Such phases
    reflect the intermediate nature of the quasi-periodic potential: in fact gapless and
    incoherent BG regions alternate with gapped incommensurate-band-insulator (IBI)
    regions, which are the quasi-periodic analog of conventional band insulators
    appearing in interacting bosonic models in a commensurate superlattice
     \cite{Rousseauetal06}.  In the limit of a deep quasiperiodic potential, the IBI phase
    can be associated to the opening of a finite gap for addition of a particle in
    any of the potential wells of the Harper potential: in this case an incommensurate
    "shell" of localised state is filled, and to start filling the next shell a gap needs to be
    overcome. This phenomenon does not occur in the case of a truly random potential,
    because the potential wells are not bounded in size, and hence the "charging energy"
    of each well can be arbitrarily small, making the system gapless at all fillings.

\subsection{Detecting disorder-induced phases}

 The above results show that a very rich showcase of quantum phases can
 be implemented by experiments with strongly interacting bosons in incommensurate
 optical superlattices. Nonetheless the fundamental question remains on how one
 can detect experimentally the occurrence of such phases and all the details of complicated
 phase diagrams. This question is ultimately related with the potential use of these
 physical systems as analog quantum simulators
  \cite{feynman82_simulators},
capable of
 providing the answer to open theoretical questions concerning models which are
 implemented literally in the experiment.

 \begin{figure}
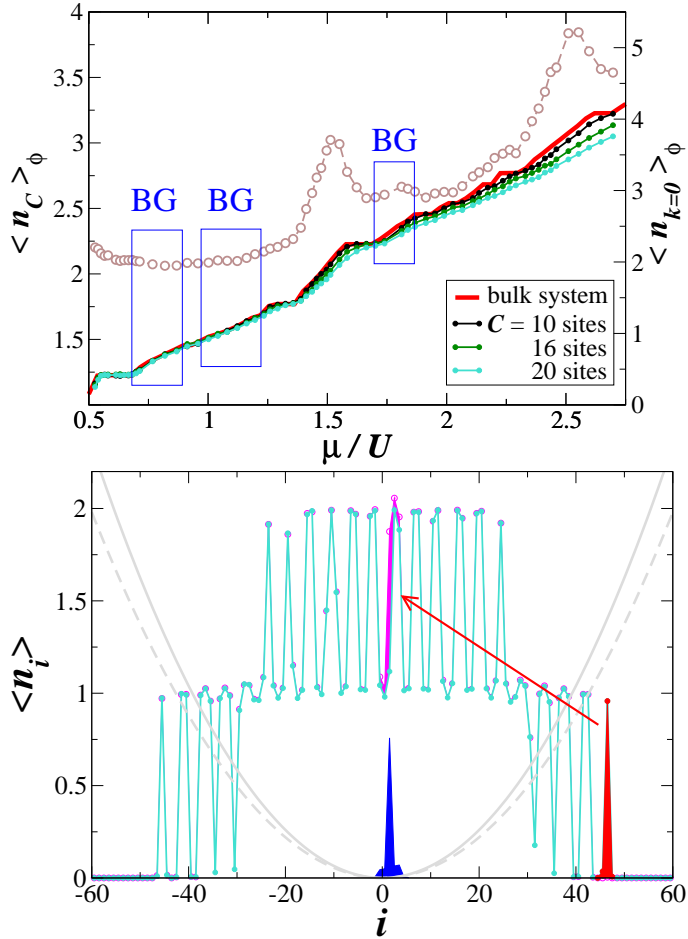

   \caption{Trap-squeezing detection of Bose glass phases. Upper panel: average central
  density and coherent fraction as a function of the chemical potential (controlled by the
  trapping potential  \cite{Roscilde09}) for a system of
  $N=100$ trapped bosons in an incommensurate superlattice with $\Delta=U/2 = 10 J$
  and $\alpha = 830/1076$; boxed regions show
  BG behaviour, namely finite compressibility without any significant increase in the coherence.
  Lower panel: trap-squeezing-induced migration of a particle from a localised state
  on the trap wings from a localised state in the trap centre. The shown data refer to a variation
  of the trapping potential $V_t$ from $0.013 J$ to $0.014 J$. }
  \label{f.squeezing}
\begin{center}
  \includegraphics[width=9cm]{Figures.CA/Trapvsbulk-1DSLavU20Nb100.eps}
  \includegraphics[width=9cm]{Figures.CA/denschange-squeezing3.eps}
\end{center}
\end{figure}

In the specific case of novel insulators induced by quasi-disordered
potentials, our recent activity has been focused on how to detect their most
striking feature, namely the simultaneous absence of a gap and of long-range
phase coherence  \cite{Roscilde09}. In particular the gapless excitations of a
BG are particle excitations, hole excitations, and particle-hole ones, and they
are probed directly by the compressibility, which is indeed finite in the BG
phase. The trapping potential confining the atoms offers the possibility of
probing the compressibility by studying the response of the system to the
increase of the trapping frequency ("trap squeezing"). Indeed such an operation
tries to push particles from the trap wings  towards the centre, where new
particles are accepted only if the local state of the system has a vanishing
gap to particle addition. Hence measuring a finite response of the average
density in the centre of the trap upon increasing the trap frequency amounts to
probing the realization of a locally compressible phase in the trap centre.

Knowing that this phase actually corresponds to a BG requires the additional
information on short-range coherence. The phase coherence of trapped cold atoms
can be measured with time-of-flight measurements, which reveal the momentum
distribution and in particular the zero-momentum component (coherent fraction):

\begin{equation}
n_{k=0} = \frac{1}{N} \sum_{ij} \langle b_i^\dagger b_j \rangle ,
\end{equation}

 where $N$ is the total particle number. In particular, if a particle is added to a BG
 phase, it will occupy a Anderson localised quasi-particle state, and hence it will
 not increase the coherent fraction of the system. Consequently a BG phase
 realised in the centre of the trap should be identified with a finite response of
 the central density to trap squeezing and on the simultaneous absence of
 response in the coherent fraction. This is indeed what it is observed in QMC
 simulations on trapped bosons subject to a variable trapping potential,
 as shown in Fig.~\ref{f.squeezing}. The realization of such a behaviour
 occurs for values of the chemical potential which correspond to
 BG behaviour of the homogeneous system, so that this type of measurement
 allows to reconstruct the phase diagram of the system {\em in absence} of
 the trap. The major requirement of this proposal is the measurement
 of the average density over a region of a linear size of $5-10 \mu$m,
 corresponding to 10-20 lattice sites.  This is indeed possible with current
 large-aperture optics  \cite{Sortaisetal07}.

\section*{Acknowledgements}

The authors would like to acknowledge the ANR support through the grants
BLANC mesoGlass and PNANO QuSpins, and the computing facilities (PSMN) of the ENS Lyon.

\bibliographystyle{plain}

\end{document}